\documentstyle[epsfig]{mn}

\begin{document}
\journal{Preprint astro-ph/9907209}
\title[The cloud-in-cloud problem for non-Gaussian density fields]
{The cloud-in-cloud problem for non-Gaussian density fields}
\author[Pedro P. Avelino and Pedro T.~P.~Viana]
{Pedro P. Avelino$^{1,2}$ and Pedro T.~P.~Viana$^{1,3}$\\ 
$^1$Centro de Astrof\'{\i}sica da Universidade do Porto,
Rua das Estrelas s/c, 4150 Porto, Portugal\\
$^2$Departamento de F\'{\i}sica da Faculdade de Ci\^{e}ncias da Universidade do Porto\\
$^3$Departamento de Matem\'{a}tica Aplicada da Faculdade de Ci\^{e}ncias da Universidade do Porto\\}
\maketitle
\begin{abstract}
The cloud-in-cloud problem is studied in the context of the extension 
to non-Gaussian density fields of the Press-Schechter approach for the calculation 
of the mass function. As an example of a non-Gaussian probability 
distribution functions (PDFs) we consider the Chi-square,  
with various degrees of freedom. We generate 
density fields in cubic boxes with periodic boundary conditions and then determine the 
number of points considered collapsed at each scale through an hierarchy of smoothing 
windows. We find that the mass function we obtain differs from that predicted using 
the Extended Press-Schechter formalism, particularly for low values of $\sigma$ 
and for those PDFs most distinct from a Gaussian.
\end{abstract}

\section{Introduction}

One of the distinctive features in the Universe is the presence of 
gravitationally collapsed structures, like galaxies, groups and clusters 
of galaxies. The distribution of masses of these structures, usually 
called the mass or multiplicity function, has been determined observationally 
\cite{ASP,HA,ECFH,Mark}, and is one of the most important characteristics of the 
Universe that proposed cosmological models attempting to explain the formation of 
structure need to reproduce. 

In order to proceed with a comparison between these observations 
and the theoretical expectations of different structure 
formation models, it is of fundamental importance to be able to 
predict with reasonable accuracy the mass function associated with the 
theoretical models. The most direct way of doing this is to perform 
N-body simulations (for a recent review see Bertschinger 1998), 
where a distribution of dark matter particles
alone, or in conjunction with gas particles, is evolved under gravity. 
However, these simulations take a considerable time to complete, being 
impracticable if a large number of structure formation models are being 
studied simultaneously. Another method for estimating the theoretical 
mass function is to use analytical approximations. Among the several that 
have been proposed, the framework put forward by Press and Schechter (1974, 
hereafter PS) has proved the most successful in reproducing the mass function 
obtained through N-body simulations, albeit the very simplified assumptions 
that are made. 

Until recently the approach proposed by Press and Schechter to estimate 
the mass function was almost always (but see Lucchin \& Matarrese 1988) 
applied in the context of structure formation models where the 
perturbations induced in the density field have a Gaussian random-phase distribution 
independently of the scale considered. This assumption is not 
only the simplest to take, but is also expected when perturbations are produced 
by an inflationary phase in the very early Universe (e.g. Liddle \& Lyth 1993). 
However, there is at present a renewed interest 
in structure formation models which predict a non-Gaussian density 
distribution, either within the context of inflation \cite{Pee99,Sal,MRS}, or as a 
result of the dynamics of topological defects \cite{Avetal98a,ABR,ACM,CHM}. 
In an attempt to compare their predictions with the observed mass 
function, particularly at the scale of galaxy clusters (Chiu, Ostriker \& Strauss 1998; 
Koyama, Soda \& Taruya 1999; Robinson, Gawiser \& Silk 1999a,b; Willick 1999) 
a particular generalization of the Press-Schechter framework has been used,
so that the assumed density field no longer needs to be Gaussian. 
This so-called Extended Press-Schechter (EPS) approach has recently been 
proved to be quite successful in reproducing the results for the mass function 
obtained from N-body simulations with non-Gaussian conditions \cite{RB}. 

The EPS method was obtained by closely following the reasoning behind the original 
PS work, in the hope that it would end up as successful in predicting the mass 
function. The fact that this seems to be the case increases our perplexity as to why 
the general PS framework works at all, given all the simplifications it entails, like 
spherical collapse and that at a given smoothing scale all the structures that form 
have equal mass. The particular issue we will study here is the so-called 
cloud-in-cloud problem, with the others addressed in forthcoming papers. This aspect 
of PS-based approaches to the calculation of the mass function has been investigated 
previously (Epstein 1983, 1984; Schaeffer \& Silk 1988; Peacock \& Heavens 1990; 
Bond et al. 1991; Jedamzik 1996; Yano, Nagashima \& Gouda 1996; Monaco 1998), 
although always in the context of Gaussian initial conditions. 

Until now the cloud-in-cloud problem within the EPS framework has been dealt with in the 
same simplified manner as in the original PS derivation. In order to determine to what 
extent such a treatment of the problem is justified, we follow the numerical approach 
laid down by Monaco (1997a,b; 1998), simulating density fields with non-Gaussian one-point 
probability distributions in cubic boxes with periodic boundary conditions.

In the first section we describe succinctly the Press-Schechter approach to the 
derivation of the mass function, and how it can be extended to accommodate non-Gaussian 
initial conditions. In the following section we discuss the cloud-in-cloud problem and 
how several authors have tried to avoid it, describing in detail the numerical 
approach we have chosen to study it. Finally, in the last two sections we present the 
results of our analysis and discuss their importance to the proposed Extended 
Press-Schechter framework.

\section{Extended Press-Schechter}

The Press-Schechter theory was originally proposed \cite{PS}, in the 
context of initial Gaussian density perturbations, as a simple 
analytical tool for predicting the mass fraction 
associated with collapsed objects with mass larger than some given 
mass thereshold $M$. This is obtained by measuring the fraction of space 
in which the evolved linear density field exceeds a given 
density contrast $\delta_{\rm c}$,
\begin{equation}
\label{PS}
F(>M)={{\Omega_{\rm m}(> M(R))}\over{\Omega_{\rm m}}}
=\int^\infty_{\delta_{\rm c}}{\cal P}(\delta) d\delta\,.
\end{equation}
For spherical collapse in an Einstein-de Sitter universe $\delta_{\rm c}$ equals 
$1.7$, being almost insensitive to a change in the assumed background cosmology 
(e.g. Eke, Cole \& Frenk 1996). In the case of Gaussian initial conditions then,  
\begin{equation}
\label{PSG}
F(>M)=\frac{1}{2}{\rm erfc}\left({{\delta_{\rm c}}\over{{\sqrt 2}\sigma(R)}}\right)\,,
\end{equation}
where ${\rm erfc}$ is the complementary error function and $\sigma(R)$ is 
the dispersion of the density field at the scale $R$. For a top-hat window, 
$M$ is related to $R$ via $M=4\pi R^3 \rho_{\rm b}/3$, with $\rho_{\rm b}$ 
being the background density. The right hand side in expression (\ref{PSG}) is usually 
multiplied by a factor of $2$, as originally suggested by Press and Schechter, so that 
\begin{equation}
\label{PSG1}
\int_{0}^{\infty}\partial F(>M)=1\,,
\end{equation}
thus taking into account the accretion of material initially present in 
regions underdense at the smoothing scale $R$. The mass function of collapsed 
objects can be obtained simply by deriving expression (\ref{PSG}) (multiplied by the 
factor $2$) with relation to $M$, and then dividing by $\rho_{\rm b}/M$. 

The original PS approach can be easily generalized for non-Gaussian 
density perturbations, simply by considering in expression (\ref{PS}) a non-Gaussian 
one-point probability distribution function (henceforth PDF) ${\cal P}(\delta)$. In order 
for all the mass in the Universe to be accounted for the expression then needs to be 
multiplied by $f=1/\int_{0}^{\infty}{\cal P}(\delta)d\delta$. This re-normalization is 
equivalent to multiplying by the factor $2$ in the Gaussian case. Surprisingly, this very 
simple extension of the PS approach to non-Gaussian initial conditions was only taken 
seriously for the first time by Chiu, Ostriker and Strauss (1997). Since then it has 
been tested with some success for a few non-Gaussian structure formation 
models against N-body simulations \cite{RB}. 

It should be noted that as long as the dispersion of the density field depends on the smoothing scale 
$R$ considered, $\sigma(R)$, the PDF for the matter density is necessarily scale-dependent, 
${\cal P}(R,\delta)$. What is usually meant by saying that such PDF is invariant with scale, is that 
the {\em reduced} distribution, ${\cal P}_{R}(\delta)={\cal P}[R,\delta/\sigma(R)]/\sigma(R)$, is 
always the same, i.e. the {\em shape} of the reduced PDF does not depend on the scale under 
consideration. For example, the simplest inflationary models predict a Gaussian PDF for the matter 
density at all scales (e.g. Liddle \& Lyth 1993), in the sense that at any scale the reduced 
form of the PDF is always equal to a Gaussian with zero mean and dispersion unity. 

In the case of alternative structure formation models one could have two other scenarios. In the 
more general one, the shape of the reduced PDF for the matter density depends on the smoothing 
scale being considered. This is indeed what is expected when the matter density distribution 
is generated, for example, through the dynamics of topological defects \cite{Avetal98b,AWS}.
The second possibility is that the reduced PDF for the matter density is scale-independent, but 
non-Gaussian, i.e. not equal to a Gaussian with zero mean and dispersion unity. Here, for 
simplicity, we will focus our study on this second scenario, with the knowledge that it can be 
easily generalized to the first one. 

\section{Solving the cloud-in-cloud problem}

\begin{figure*}
\centering
\epsfysize=4.8cm \epsfbox{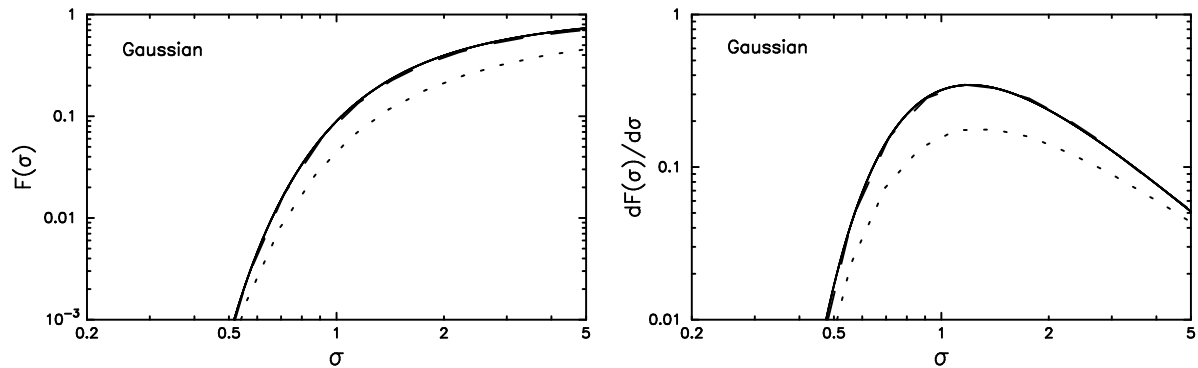}
\caption[Figure1]{For the reduced Gaussian PDF it is shown on the left $F(\sigma)$, and 
on the right its derivative with respect to $\sigma$. The solid lines correspond to 
the EPS prediction, while the dashed and dotted lines to the numerical results, respectively 
in the case of sharp-k and gaussian smoothing.}
\end{figure*}

As we have previously shown, there is a fundamental difficulty with the 
normalization of expression (\ref{PS}). We have seen that in the Gaussian 
case a factor of $2$ had to be introduced to correctly take into account 
the fact that material in initially underdense regions at some smoothing scale R 
is eventually accreted and incorporated into the collapsed objects that form 
from the initially overdense regions at that scale. In other words, the factor $2$ 
accounts for the material that although not in the regions predicted to collapse at 
the smoothing scale $R$, will nevertheless become part of collapsed objects 
associated with scales larger than $R$. Only by smoothing the density field at these 
scales would this material count as collapsed, but clearly this should happen in 
the first place when $F(>M(R))$ is calculated. This is the so-called 
cloud-in-cloud problem in the PS approach. 

Several authors have tried to find a more satisfactory solution for the cloud-in-cloud 
problem than just multiplying expression (\ref{PS}) by $2$, as proposed by Press 
and Schechter. The first to approach the problem were Epstein (1983, 1984) 
and Schaeffer \& Silk (1988). But it was only with the work of Peacock and Heavens (1990) and 
Bond et al. (1991) that a comprehensive framework was put in place to study the cloud-in-cloud 
problem and its ramifications. The assumptions behind the two approaches are very similar, 
with the later using a more formal line of reasoning based on the theory of excursion sets. 
The method we will use here to study the cloud-in-cloud problem within the EPS 
framework was first used by Monaco (1997a,b; 1998) 
for the case of Gaussian density fields. It is basically a numerical implementation of the 
approach pioneered by Peacock and Heavens (1990) and Bond et al. (1991). Density fields 
with an assumed PDF, and which only differ in the scale at which the 
smoothing is applied, are generated in cubic grids. Starting from the largest 
scale, trajectories for all the points in the density field are then constructed with 
the values for the density contrast recorded at each smoothing scale. In the Gaussian case, 
and if the smoothing is performed with a sharp-k window, the trajectories followed by the 
points are Brownian random walks. In this very particular example, the number of points 
which exceed some density contrast at any smoothing scale is equal to the number of points 
which though at such scale do not exceed the assumed density contrast, at some other 
{\em larger} scale do. In this case the re-normalization factor one needs to multiply 
expression (\ref{PS}) with is exactly $2$, as Press and Schechter initially proposed. 
This was formally proved by Bond et al. (1991) using excursion set theory. Unfortunately, 
it is the only instance when their approach can be used to solve the cloud-in-cloud problem. 
For any other smoothing window or probability distribution the random walk characteristic 
disappears, and the problem becomes analytically intractable. In these cases, either one 
uses the method proposed by Peacock and Heavens (1990), and expanded in Monaco (1997a,b; 1998), 
which also has its limitations, or the numerical approach considered here.  

\begin{figure*}
\centering
\epsfysize=11.5cm \epsfbox{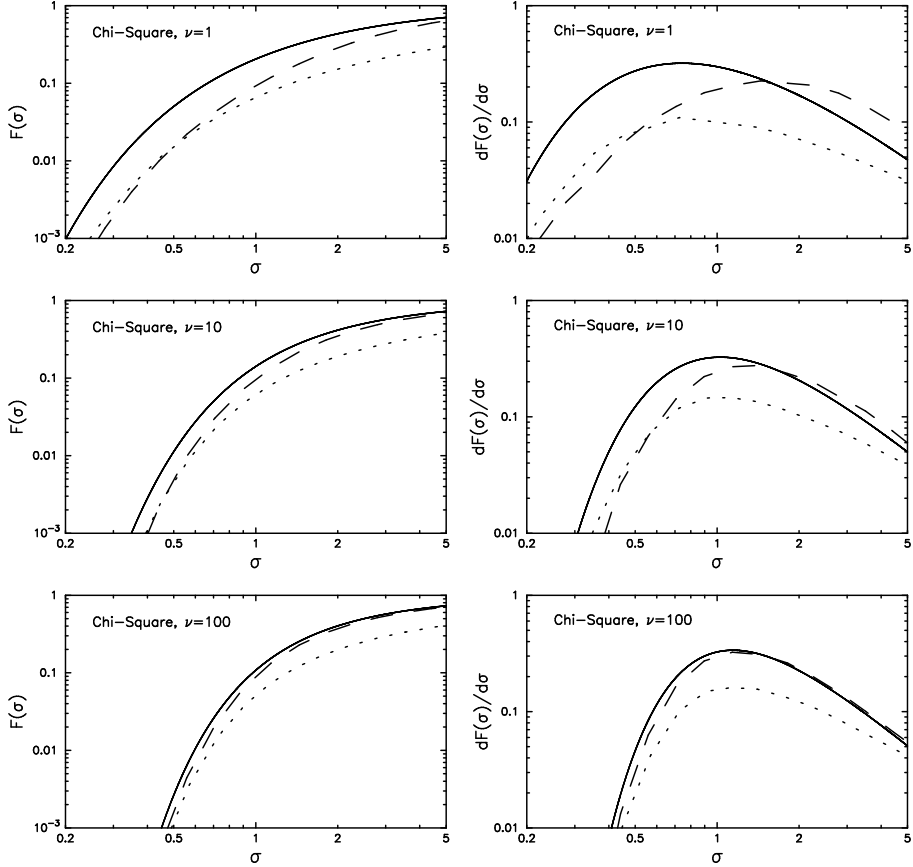}
\caption[Figure2]{The same as in Figure 1, but for the three reduced Chi-square
PDFs considered in the paper.}
\end{figure*}

\section{Results}

Our numerical realizations of density fields were performed in cubic grids with $64^3$ 
points. They were generated in the same manner as the density fields used to set initial 
conditions in N-body simulations (see e.g. Bertschinger \& Gelb 1991; Klypin \& Holtzman 1997). 
We considered one example of a scale-independent reduced non-Gaussian PDF for the density contrast, 
the Chi-square, the PDF having been shifted so that its mean is zero. Such a PDF is expected in 
certain models of structure formation, involving either isocurvature density 
perturbations generated during an inflationary period in the very early Universe \cite{Pee99} or 
cosmic string seeded perturbations \cite{Avetal98b,AWS}. We also considered the Gaussian case for comparison. 

The Chi-square has the added attractive feature that the reduced PDF 
remains approximately the same when smoothed on a variety of length scales using either 
gaussian (GAU) or top-hat (TH) windows (e.g. White 1999). The reduced PDF that results from 
the smoothing becomes increasingly different from the original as the number of degrees of freedom, 
$\nu$, gets smaller, with the most important departures relatively to the original reduced PDF shape being 
a decrease in the probability of $\delta/\sigma$ taking values around zero, i.e. the mean, 
and the appearance of a non-zero probability of $\delta/\sigma$ taking values just outside the cut-off, equal 
to $-\nu/\sqrt{2\nu}$. Fortunately, these differences only become noticeable for $\nu$ smaller than about 10. 
However, when the smoothing is performed using a sharp-k (SK) window, the departure of the smoothed PDF 
from the original increases considerably. Now, as soon as $\nu$ becomes less than about 100 the difference 
starts to show. Nevertheless, we will be only interested in the fraction of the reduced PDF 
that lies above $\delta_{\rm c}/\sigma$, and this part of the distribution is little changed by smoothing, 
even in the case of a SK window.

With the above in mind, for different $\nu$ values we assumed that the power 
spectrum associated with the density contrast was a power-law, 
${\cal P}_{\delta}(k)\propto k^{n}$, where $k$ is the wavenumber and $n=-2,-1,0$. We 
generated several sets of density fields for each combination of $\nu$ and 
$n$ values, such that all realizations within each set only differed in the 
smoothing scale applied. The three smoothing windows mentioned above were considered: sharp-k, 
top-hat, and a gaussian. However, we will only show the results for SK and GAU 
smoothing, given that those for TH smoothing turn out very similar to the GAU ones. Also, 
we opted for showing just the results obtained for $n=-2$, which is closest to the slope of 
the matter power spectrum on the scale of galaxy clusters \cite{Mark}, as they are basically 
indistinguishable from the results for the two other spectral indexes, $n=-1$ and $n=0$. 

On the left panel of Fig. 1 we show, for the reduced Gaussian PDF, the fraction of mass 
collapsed above a certain smoothing scale as a function of the value of the dispersion of 
the density field at that scale, $F(\sigma)$, calculated using the method presented 
here and the EPS approach. The same quantities are shown on the left panel of Fig. 2 
for a reduced Chi-square PDF with $\nu=1$, $\nu=10$, $\nu=100$ 
(note that the $\nu=\infty$ case is equivalent to a reduced Gaussian PDF).

On the right panels of Fig. 1 and Fig. 2 we show $dF(\sigma)/d\sigma$ instead. 
The theoretical mass function can be easily obtained by multiplying $dF(\sigma)/d\sigma$  
by $d\sigma/dM$ and dividing by $\rho_{\rm b}/M$. In the case of power-law spectra as 
the ones we consider here, $\sigma(M)\propto M^{-(n+3)/6}$. 

In all cases the thereshold for the density contrast above which 
we assume a field point to be collapsed is $\delta_{\rm c}=1.7$, which is equivalent to 
assuming spherical collapse in an Einstein-de Sitter cosmology. The results presented 
can be easily generalized for any other thereshold, $\delta_{\rm c}'$, by making the 
identification $F'(\sigma)\equiv F(\sigma*\delta_{\rm c}/\delta_{\rm c}')$. 

In the case of SK smoothing and in the limit of a reduced Gaussian PDF, the EPS 
approach (which then simply reduces to PS) correctly predicts $F(\sigma)$, 
and thus $dF(\sigma)/d\sigma$, as expected. The famous factor of $2$ in 
the normalization is recovered. As the number of degrees of freedom decreases the assumed reduced 
PDF starts to differ from a reduced Gaussian, and the EPS prediction increasingly overestimates the 
numerically determined collapsed mass fraction and mass function. 
In the case of the GAU window (as mentioned before the 
results for TH smoothing are very similar), the numerical 
results deviate from those predicted through the EPS approach even when assuming a 
Gaussian PDF. For this particular PDF, we find that for small values of $\sigma$ 
(i.e. large mass scales), both the predicted collapsed mass fraction and the mass function 
seem to approach the result one would obtain if the PS approach was used without 
the normalization factor of 2. The same conclusion had already been reached 
analytically by Peacock and Heavens (1990), being numerically confirmed 
by Bond et al. (1991) (see also Monaco 1997b). 

\section{Conclusions}

We have shown that the EPS approach does not correctly take into account the existence of 
regions, which though at some smoothing scale are unable to pass a certain collapse 
thereshold are nevertheless able to do it at larger scales, in the estimation of the mass 
function. This problem was already known to exist within the PS framework, except when the 
smoothing of the density field was performed using a sharp-k window. Now, we have found that even 
when this window is used, the EPS approach cannot adequately solve the cloud-in-cloud problem. 

The mass function predicted by the EPS approach deviates most from the numerical results 
for small values of $\sigma$, which correspond to the large mass end in models where structure 
builds up hierarchically. The only N-body simulations that have been used to check whether 
the EPS approach provides a good fit to the mass function were limited in range to values 
of $\sigma$ larger than around 0.5, except for a couple of points with relatively large error 
bars, and at the limit of statistical significance \cite{RB}. For these $\sigma$ values, the EPS 
prediction for the mass function has not yet entered the regime where it differs significantly 
from the numerical results obtained in this paper, particularly when one considers sharp-k 
smoothing. It is therefore not surprising that the N-body results seem to vindicate the EPS 
approach. However, our analysis should throw a note of caution. In the regime where the mass 
function is defined essentially through the abundance of rare density peaks, the EPS approach may 
not be reliable when one is dealing with strongly non-Gaussian PDFs. This may affect some recent 
conclusions regarding the gaussianity of the density field on large scales, drawn from the 
abundance of rich galaxy clusters at different redshifts and their present-day correlation length 
(Chiu, Ostriker \& Strauss 1998; Koyama, Soda \& Taruya 1999; 
Robinson, Gawiser \& Silk 1999a,b; Willick 1999). 
It would be very interesting if larger N-body simulations could be carried out, able to 
extend the mass function further into the low-$\sigma$, rare events regime. 

The mass functions determined in this paper can still be further improved. 
Here we focused our attention on the cloud-in-cloud problem. Other issues were left untouched, 
more importantly, possible deviations from spherical collapse and the relation between smoothing 
radius versus the mass of the structures identified after each smoothing. We are presently 
looking at these issues.

\section{Acknowledgments}

P.P.A. and P.T.P.V. are supported through the PRAXIS XXI program of FCT.

\end{document}